# Privacy vs National Security

Tajdar Jawaid

*MS Cybersecurity, University of Dallas, TX, USA*

**Abstract :** There are growing concerns and anxiety about privacy among general public specially after the revelations of former NSA contractor and whistle-blowers like Edward Snowden and others. While privacy is the fundamental concept of being human, the growing tug-of-war between individual's privacy and freedom vs national security has renewed the concerns of where the fine balance should lie between the two. For the first time in history the technological advancement has made the mass data gathering, analysis and storage a financially and technologically feasible option for the governments and private businesses. This has led to the growing interest of governments and security agencies around the globe to develop sophisticated algorithms using the power of Big-Data, Machine-Learning and Artificial Intelligence. The technology has enabled governments and private businesses to collect and store thousands of data points on every individual, which has put individual's privacy under constant threat. This article analyses the individual's privacy concepts and its perceived link with national security. The article will also discuss the various aspects of privacy and national-security, arguments of both sides and where a boundary should be drawn between privacy and national-security.

**Keywords:** *Privacy, Human Right, National Security, Terrorism, Counterterrorism, Profiling, Cataloguing, Personal Data, Mass-Surveillance, Cybersecurity, Big-brother, Machine-Learning, Big-Data, AI, FISA.*

## I. INTRODUCTION

There was a time when surveillance on anyone be it is an individual, organisation or enemy state was considered to be a very difficult, sophisticated, time and resource consuming operation. There were numerous laws, financial constraints, and technological hurdles. For instance if a surveillance is required on an individual the security or law enforcement agencies has to create proper justifications to get the necessary warrants and relevant court permissions, allocate massive resources, break-in into the premises of the target to install bugs for surveillance, physical spies are needed on targets to monitor and photograph all movements and interactions, wiretap phones, listen conversations, collect evidences and so on. All these constraints on one hand make the job of surveillance and intelligence gathering difficult, but on the other they ensure the greater privacy to the individuals and deter the governments and its agencies to initiate these activities without a justifiable cause. In past few years, the advancement in technology has made massive paradigm shift in surveillance. Technology has now made the intelligence gathering not only financially viable but very attractive options for governments around the globe. Since 9/11 and the resulting war on terror, there was a strong desire among the security and law-enforcement communities to find out ways to gather the intelligence at massive scale to avoid such incidents in future. This desire turns into a requirement of security and intelligence agencies that has given the birth of the concept of mass-surveillance. Mass-surveillance programs has enabled governments and its intelligence agencies to monitor large set of population without their knowledge and consent. The rise of social-media, internet connected smart devices, IoT (internet of things) also have deep impact on individual's privacy. These concerns result in debates around what actually is considered to be a private information in this time and age. The next section will focus on what actually privacy is in current scenario, why it is important and why it is necessary to be protected.

## II. BACKGROUND ON WHY PRIVACY MATTERS

Every human being has things which they wanted to keep very personal. It does not mean that it has be to something illicit, unlawful or criminal in nature. In fact, most of the time these are very innocent, totally harmless personal information in terms of religious, political or social views, some very personal choices, desires and feelings, sexual orientation, relationships, health related information and son on. Still individuals are not comfortable to share it with anyone, even to their closest ones. This is the actual essence of privacy, disclosing these kinds of information is a personal choice and a fundamental human right. Privacy has been recognised as the fundamental human right under the United Nations declaration on Human rights (UDHR) article 12 [1], which states, *"No one shall be subjected to arbitrary interference with his privacy, family, home or correspondence, nor to attacks upon his honour and reputation. Everyone has the right to the protection of the law against such interference or attacks.",* which was reaffirmed through the United Nation's International Covenant on Civil and Political Rights (ICCPR) article 17:1[2]. Accordingly, to the UN declaration and civil and political rights covenants the law should protect and provide appropriate measures to ensure the protection of every individual's privacy.

Privacy ensures the individual's right to the freedom of speech and expression, it protects from race/religious/political/sexual persecution, provided the right to freedom and choice to share or hide personal information from others. Privacy is a state of not being watched or disturbed without our knowledge and consent. Right to privacy ensures that we are free from state surveillance, free to have our own unique thoughts and views, free from being just a number, free from profiling and cataloguing based on various character traits, free to protest, free to vote, free to think, free be left alone and so on so forth. Privacy as some view is what keep us separate from zoo animals who are continuously being watched and filmed without their will and consent. Privacy is very basic human instinct and is recognised as a fundamental human right, it gives the confidence of being in-possession of our own personal information, thoughts, views and opinions without being judged. These basic concepts of privacy are now considered to be under threat as per organizations and movements like privacy international [3]. To protect this fundamental right there are multiple laws enacted which will be discussed in next section.

### III. PRIVACY LAWS AND REGULATIONS

The privacy related issues and concerns in Europe goes back as far as 1970's and 80's, as government agencies and privates' businesses started to collect and store customer data. As a result, in Europe common protection system implemented which was followed by EU Data Protection Directives in the 1990s (directive 95/46/EC) [4]. The privacy concerns increase with the growing governments and private business interest in gathering and storing personal information. To address these concerns in Europe the EU Data Protection directives of 1990 is replaced by EU-GDPR (European Union General Data Protection Regulation) in 2016 [5].

In United States the fourth amendment considered to be basis of most privacy laws, which states that, *"The right of the people to be secure in their persons, houses, papers, and effects, against unreasonable searches and seizures, shall not be violated, and no Warrants shall issue, but upon probable cause, supported by oath or affirmation, and particularly describing the place to be searched, and the persons or things to be seized"*[6]. In US multiple federal and state specific laws were enacted to ensure the privacy rights of an individual. Some of these laws in U.S. are, Children's Online Privacy Protection Act (COPPA) of 1998, The California Online Privacy Protection Act (CalOPPA), Electronic Communications Privacy Act (ECPA) of 1986, Communications Assistance for Law Enforcement Act (CALEA) of 1994, Health Insurance Portability and Accountability Act (HIPAA) of 1996, Health Information Technology for Economic and Clinical Health Act (HITECH) of 2009, Data Breach Notification Laws, Family Educational Rights and Privacy Act of 1974 and Identity Theft and Assumption Deterrence Act of 1998 are the few. These and other privacy related laws and regulations provided the required framework for governments and private organisations to implement stringent security to ensure the privacy of an individual. The control over privacy is getting weaker even though the privacy rights grown stronger. This is due to the fact that personal data are now much more exposed and easily available, which was not the case few years back.

It is important to understand what constitute personal data and what comes under privacy. At very high-level any information that defines and uniquely identifies a person normally classified as personal data and should be protected by privacy laws and regulations. This means in an ideal world any information falls under personal data should not be harvested without the knowledge and consent of the data subject. What comes under private personal data and falls under privacy right will be discussed in the next section.

### IV. WHAT ACTUALLY COMES UNDER PERSONAL DATA

Technically anything considered to be private related to personal data which a person is not willing to share with the world. But according to European General Data Protection Regulation (GDPR) of 2016, the personal data means any information about a natural person (living person) which identify a natural person directly or indirectly, in particular by reference to an identifier such as a name, an identification number, location data, an online identifier or to one or more factors specific to the physical, physiological, genetic, mental, economic, cultural or social identity of that natural person. Together with this the article 9 of GDPR prohibits the processing of following categories of personal data e.g. race, ethnic origin, political opinions, religion, philosophical beliefs, trade union membership, genetic data, biometric data, health data, concerning a natural person's sex life, sexual orientation etc. without an explicit consent of data subject. The GPDR has provided a clear and concise definition of the personal data or PII data, for which organisations has to have legitimate business and lawful purposes and data subject consent to process and store. Whereas the article 12 to 17 of the GDPR gives multiple rights to data subject related with their personal data, such as right of access, right to be forgotten, right to restrict processing, right to data portability, right to object to processing [5].

In digital and internet context, the personal data includes but not limited to person's name, gender, race, beliefs, address, photographs, videos, friends and family members details, health related information, physiological information, biometrics, phone numbers, IP address, location data like GPS coordinates or Cell-ID locations, and so on. Any digital trace and information which uniquely identified a person is considered to fall under personal data [5].

The next section will discuss the link between individual privacy with national security.

## V. THE INDIVIDUAL PRIVACY VS NATIONAL SECURITY

In 1984 George Orwell wrote a novel called *"Big-Brother"*. The basic premise of the novel was that every citizen will be continuously watched, listened, tracked, profiled and catalogued by the government. This fictional work has envisioned the environment where there was no concept of citizen's privacy at all [7]. In the aftermath of September 11, 2001, the United States government passes the USA Patriot Act on October 26, 2001 which has given new dimensions to the privacy of an individual by linking it to the national security [8]. This perceived link between national-security and citizen's privacy trend has been followed by multiple governments around the globe. The United Kingdom has introduced one of its own bills in 2016 which gives immense powers of surveillance to its security agencies with the bill called Investigatory Powers Act [9]. These kind of counterterrorism laws has given the enhanced power to the security and intelligence agencies to allow surveillance by all possible means e.g. digital communication monitoring which includes (phones, emails, text messages etc.), investigating any suspect without tipping off, obtain business records, cross-border information and intelligence sharing, obtain search warrant anywhere, monitor electronic trespassers, enhances the punishments etc. These laws aim to equip the intelligence and security agencies with the necessary tools to intercept and obstruct terrorism. These counterterrorism measures have result in several covert indiscriminate mass-surveillance programs from governments around their citizens. However, the full extent of these programs and their invasiveness to citizens privacy first surfaced after the revelations made by former NSA whistle-blower.

### A. The Snowden's Leaks: "A Case Study"

The startling revelations by former NSA contractor like Edward J. Snowdon has raises a new and legitimate concerns around the privacy invasion in the name of national security. The scenario which was presented in Orwell's novel that a state act like *"Big-Brother"* by initiating the mass indiscriminate surveillance, tracking, profiling and cataloguing of its citizens seems to become a reality by Snowden's leak [10].

*The mass-indiscriminate Surveillance:* The Snowden leaks raises the concerns over a mass-indiscriminate surveillance of all citizen communication over the internet. This includes obtaining of all phone records, emails and text messages from service providers, all social-media posts, blogs and vlogs etc. The leaks also reveal that in many cases FISA process has been by-passed for the collection of citizen's private information [10].

*Gather and Store all communication:* The leaks also discuss the creation of big datacentres with virtually unlimited or expandable storage capacities. They are created to gather and store all digital communications over the internet for later analysis [10].

*The Meta-Data:* The government stance is that they only collect the meta-data about communication, which does not contain the actual communication itself. But Snowden suggest that it is not entirely true, there are programs which gather and store actual communications e.g. phone calls, emails, text message etc. [10].

*The Prism Program:* The Snowden's leak has revealed the details about the Prism Program. The prism program provides the most detailed search capability on any individual through the access on social media platform like Apple, Google, Amazon, Facebook, Twitter, etc. These social media platforms provided the real-time search API's to NSA. These API's then combined into a single search interface, which can pull all the information about any specific person on various search criteria from the servers of the above companies. This includes data from all social-media posts, connection details, friends and family details, emails, texts messages, photographs, internet searches in fact anything or everything stored on their platforms about an individual [10].

*Zero-day Exploits (the backdoors):* Zero-day exploits are unknown, bugs and backdoors. Sometime deliberately created specifically for surveillance purpose in mobile, desktops operating systems, applications and software. Through these zero-day exploits any internet connected device camera, mic can be remotely activated without the device owner/user noticing it or the mobile applications track gather and transport personal data like phone records, contact details, emails, text messages on the servers without ever being noticed. Through these exploits a person effectively can be watched or heard, tracked through their device's GPS all without their knowledge and consent. Although all big names like Apple and Google are opposed to this request from security and intelligence agencies. But according to Snowden's leak that was also a part of prism program [10].

### B. A Problem without a clear Solution

The terrorist incidents like 9/11 considered to be the failure of security and intelligence agencies. Incidents like this has been the primary motivation behind the laws like the USA Patriot Act and British government's Investigatory Powers Act. These kinds of laws indicate there is clear shift how governments see the privacy of their citizens. Privacy that once considered the basic human right is now on direct collision path with national security.

The primary responsibility of any state is to protect the live and livelihood of their citizens, maintain law and order and protect its national infrastructure and interests. Therefore, in order to achieve these objectives, the security agencies have to equip with necessary tools and technologies supported by the law to gather the intelligence which will help prevent the repetition of incidents like September 11, 2001.

There are also debates about the trade-offs between security and privacy. Too much privacy may hinder the ability of security agencies to gather required intelligence on targeted individuals. Also, too many prerequisites like gather evidence to obtain necessary court orders may put security at compromise. On the other hand, unlimited and unrestricted powers to invade citizen's privacy by government agencies in the name of national-security may increase the risk of power misuse which increases the trust deficit between general-public and governments.

This is still a debatable subject, as to what level of privacy invasion and compromises are acceptable to achieve security objectives. What kind of powers should be vested to intelligence communities to curb the next big attack. Are there any other counterterrorism alternatives and solutions which could give the required level of information to intelligence communities to ensure security without invading the privacy of the individuals. Should governments wait for any incident to be happened before they allow the security and intelligence agencies to be able to do the surveillance on the suspected individuals. An analysis of recent terror related incidents reveals that the majority of the suspected terrorist carried out the attacks are already known to the law-enforcement agencies, and in some instances the red-flags were already raised against them, still those individual successfully carried out the attacks [16]. This in itself raises questions around mass-surveillance programs effectiveness, their lack of ability to provide the actionable intelligence etc. These and many similar questions are part of the wider debate on this subject, without a clear-cut and agreeable solution to date.

*The Increase in terrorism:* Since the war-on-terror started post 9/11, which resulted in various mass-surveillance programs, the actual number of terrorism related incidents has been increased globally.

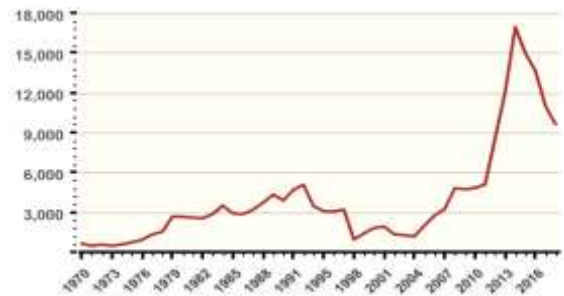

**Figure 1: University of Maryland (2020). Global Terrorism Database [14]**

Figure 1 of University of Maryland, Global Terrorism database shows record number of increases in global terrorism related incidents since 9/11 [14]. This is an indication that these programs are not as useful as it is perceived in the view of intelligence communities, at least not in their current shape and form to combat terrorism.

*The general-public trust and opposition:* Since Snowden's leaks the approval ratings from public has been decreased. In a survey in 2015 conducted by PEW research centre 52% Americans were concerned about government surveillance of Americans personal data and communication [15].

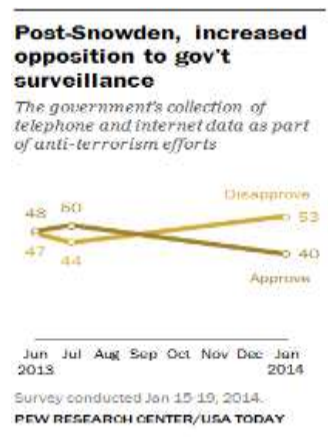

**Figure 2: PEW Research Jan 15-19, 2014 [15].**

The PEW research in 2014 shows that post-Snowden the opposition to government surveillance has increased [15].

### C. The False Trade-off between Privacy vs. National Security

There is a general perspective that there are trade-offs between privacy and security. However, these perspectives often based on arguments which are not fully supported by substantial facts and figures.

*Good People vs Bad People View:* There are views that sometimes enforces by the influencers like the technology giant Google, its CEO Eric Schmidt once said *"If you have something you don't want anyone to know, May be you shouldn't be doing it"* [11]. This particular mentality shows that the privacy is a concern for only bad people, who are doing bad things e.g. involved in criminal activities, terrorism etc. They completely ignore the importance of privacy or the fact that the privacy is not about being good or bad, but a basic human right.

*Nothing to hide, Nothing to Fear:* This is again on the similar lines of good and bad people view. This view again argues that essentially the reason to hide personal data from government is because of the fear of something wrong in it. Again, people with these views completely ignores what the privacy actually means [12].

*The cost of Security is Privacy:* If the next 9/11 can be avoided by giving up my privacy, I am up for it. Again, the cost of security does not have to be paid by giving up the fundamental human right to privacy.

**Civil liberties and anti-terrorism policies**

| Percent who favor each as a measure to curb terrorism | Sept 2001 % | Aug 2002 % | Dec 2006 % | Aug 2011 % |
|---|---|---|---|---|
| Requiring that all citizens carry a national ID card at all times | 70 | 59 | 57 | 57 |
| Extra airport checks on passengers who appear to be of Middle-Eastern descent | -- | 59 | 57 | 53 |
| Government monitoring credit card purchases | -- | 43 | 42 | 42 |
| Government monitoring personal phone calls and emails | -- | 33 | 34 | 29 |

**Figure 1: PEW Research Centre Aug 17-21, 2011 [15]**

Pew research in 2011 on the tenth anniversary of 9/11 reveals that 54% Americans were not in favour of giving up civil liberties as a method to curb terrorism [15].

The next section will discuss where the possible boundary can be drawn between Privacy vs National Security.

## VI. PRIVACY VS. NATIONAL SECURITY, WHERE SHOULD BE THE BOUNDARY?

There may be trade-offs between national security and privacy, but they should be based on the facts and figures. There should not be a complete blind eye by governments on the powers vested to its security agencies in the name of national security. There should be a proper system of check and balances needed to be developed supported by the law in order to avoid the misuse these powers.

*The Internet and Digital Communication:* After September 11, the NSA has initiated a domestic surveillance program known as *"President's Surveillance Program"*. The program enforces all communication and internet service providers to provide all communication and call detail records of all their customers. NSA also recorded the emails of all customers of these telecommunication providers [13]. This trend has been followed by multiple governments around the globe. In fact, these internet and telecom companies have developed search portals which are made available to the government agencies which can be used to search any communication details happened over their network in real-time. Again, this level of information access is key in intercepting, tracing and locating a suspect before they carry out their attacks. A proper government oversight is needed to address the privacy and misuse concerns.

*The Encrypted Internet Communication:* The security and intelligence agencies view the encrypted internet communication technologies as a hinderance in their work. Which in many cases may be a legitimate concern. The counter argument here is providing the cryptographic keys to agencies could make the whole internet communication unreliable and unsafe. The access to the cryptographic keys to government agencies could be a prime target for malicious hackers, put internet security on risk, compromise the human right to privacy and undermine the rule of law. These kinds of law enforcement access to encrypted communication should approached with caution [17].

*Video Surveillance:* In today's world close circuit tv and security cameras are installed in public and private place in all cities and towns, shops, bars, businesses, in buses, trains and every step of the way. Every movement of a person is recorded and watched. This video surveillance by definition is also an invasion in privacy, but it is an acceptable norm around the world. These video surveillance technologies are now increasingly used with biometric technologies such as facial recognition which can be used to profile, catalogue and identify an individual. These technologies are essential to provide security and deterrence but at the same time raises the concern over privacy invasion. The use of video surveillance technology when clubbed with biometrics such as facial recognition and related algorithms, should have a proper law oversight to avoid any misuse of these technologies.

*Big-Data Analytics, Machine-Learning and Artificial Intelligence:* Big-data technologies has revolutionized the way data has been studied. Big data has enabled the systematic analysis of very large and complex datasets to extract the meaningful information through various algorithms. The algorithms like pattern recognition and pattern matching, identify trends and associations among seemingly unrelated and very large and complex datasets. The intelligence and law enforcement agencies use big-data technologies and algorithms on the citizen personal data which they collected from different sources like internet communication service providers, social media platforms and various other means. The big-data analytics together with Artificial Intelligence and Machine learning technologies, are used to improve the prediction on the human behaviour through the process called psychographic analysis. It is a known fact that majority of terrorists leaves digital traces while communicating, planning and interacting over the internet. These technologies can be used to study, analyse and predict the behaviour of these individuals in order to combat the terrorism and security threats. But at the same time a misuse of these technologies can have devasting impact like it was revealed in the Cambridge Analytica scandal. The scandal reveals how these technologies has been used to influence the whole election processes in various countries [18]. Further research are required to develop or refine the existing big-data algorithms to help intelligence agencies with the appropriate oversights to avoid unnecessary privacy invasion of citizens supported by the laws to curb the misuse of these technologies.

***Store everything for later analysis:*** As revealed in the Snowden case study the governments and private business to create big data centres with virtually unlimited data storage capabilities. This result in storing all internet communication for unlimited period of time, to be analysed later [10]. The blanket application of gathering and dumping everything by security agencies into these big datacentres should be scrutinized by governments and relevant watchdogs. There is always a risk if a malicious hacker gets their hands on any such datastores what level of havoc they can pose to individuals, corporations and governments.

***The FISA and Investigatory Powers Act:*** The U.S government has provided the lawful backing to the security agencies through Foreign Intelligence Surveillance Courts (FISA). The FISA courts in USA ensures that the security and law enforcement agencies can only do the surveillance on American citizens and permanent residence through a proper court orders when there is a probable cause or a legitimate security concern [19]. Similarly, in UK the Investigatory Powers Act gives the enhanced powers to the law-enforcement agencies to obtain communications and data about communications and other digital traces of an individual [9].

Though the FISA courts in USA and Investigatory Power Courts in UK are enacted to justify every surveillance needs through probable cause. But according to Snowden's leaks the security agencies in majority if not in all cases completely bypass these processes [10][18], or these courts passes orders in favour of law-enforcement and security agencies without due scrutiny to ascertain the probable cause [19].

While in some cases understandably the intelligence agencies may not have enough time at their disposal to go through the due process, but at the sometime a complete blind eye may compromise the privacy and may be counterproductive to achieve security objectives. Hence it is necessary that these legal and lawful processes must be refined, strengthen and properly enforced.

## VII. CONCLUSIONS

The national security and privacy are deeply interconnected topics. Though the technological advancement for the first time in history has enabled mass-surveillance a viable technological and financial option for governments, but it has also raised concerns over the right to privacy. As privacy awareness are increasing, the privacy laws and regulations are getting tougher, but the control over privacy is getting weaker. Privacy is the fundamental human right and must be protected. The debate between privacy and national security is complex as both sides have the compelling arguments. There may be trade-offs between security and privacy, still a proper government oversight is necessary to avoid mass indiscriminate citizen surveillance and to put a proper check and balances over the powers given to the security and intelligence agencies.